\documentclass[reprint,showpacs,preprintnumbers,amsmath,amssymb,onecolumn,superscriptaddress,showkeys]{revtex4-1}
\usepackage[dvipdfmx]{graphicx} 
\def\ch{{\cal H}}
\def\cp{{\cal P}}
\def\cq{{\cal Q}}

\def\cv{{\cal V}}
\def\el{{\cal L}}

\def\hbar{\mathchar'26\mkern -9muh}

\def\v1{\mathbf{1}}

\usepackage{color}
\usepackage[pagebackref=false]{hyperref} 
\hypersetup{
    bookmarks=true,         
    unicode=false,          
    pdftoolbar=true,        
    pdfmenubar=true,        
    pdffitwindow=false,     
    pdfstartview={FitH},    
    pdfkeywords={keyword1} {key2} {key3}, 
    pdfnewwindow=true,      
    colorlinks=true,       
    linkcolor=red,          
    citecolor=blue,        
    filecolor=magenta,      
    urlcolor=cyan,           
}
\definecolor{amethyst}{rgb}{0.6, 0.4, 0.8}

\begin{document}

%
%
\title{Environmental engineering for quantum energy transport}  
\author{Chikako Uchiyama}
\email{hchikako@yamanashi.ac.jp}
\affiliation{National Institute of Informatics, 2-1-2 Hitotsubashi, Chiyoda-ku, Tokyo 101-8430, Japan}
\affiliation{Graduate School of Interdisciplinary Research, University of Yamanashi, 4-3-11, Takeda, Kofu, Yamanashi 400-8511, Japan}
\author{William J. Munro}
\affiliation{NTT Basic Research Laboratories, NTT Corporation, 3-1 Morinosato-Wakamiya, Atsugi, Kanagawa 243-0198, Japan}
\affiliation{Research Center for Theoretical Quantum Physics, NTT Corporation, 3-1 Morinosato-Wakamiya, Atsugi, Kanagawa 243-0198, Japan}
\affiliation{National Institute of Informatics, 2-1-2 Hitotsubashi, Chiyoda-ku, Tokyo 101-8430, Japan}
\author{Kae Nemoto}
\affiliation{National Institute of Informatics, 2-1-2 Hitotsubashi, Chiyoda-ku, Tokyo 101-8430, Japan}
\date{\today} 
\begin{abstract}
Transport phenomena are ubiquitous throughout the science, engineering and technology disciplines as it concerns energy, mass, charge and information exchange between systems.  In particular, energy transport 
in the nanoscale regime has attracted significant attention within the physical science community due to its potential to explain complex phenomena like the electronic energy transfer in molecular crystals or the Fenna-Matthews-Olson (FMO) / light harvesting complexes in photosynthetic bacteria with long time coherences.  Energy transport in these systems is highly affected by environmental noise but surprisingly not always in a detrimental way. It was recently found that situations exist where noise actually enhances the transport phenomena. Such noise can take many forms, but can be characterised in three basic behaviours: quantum, correlation in time, or space. All have been shown potential to offer an energy transport enhancement. The focus of this work is on quantum transport caused by stochastic environment with spatio-temporal correlation.  We consider a multi-site nearest neighbour interaction model with pure dephasing environmental noise with spatio-temporal correlation and show how an accelerated rate for the energy transfer results especially under negative spatial correlation (anti-correlation). Spatial anti-correlation provides another control parameter to help one establish the most efficient transfer of energy and may provide new insights into the working of exciton transport in photosynthetic complexes.  Further the usage of spatio-temporal correlated noise may be a beneficial resource for efficient transport in large scale quantum networks. 
\end{abstract}
\maketitle

\noindent {\textsf{\bf Introduction}}\\
Energy transport is a fundamental primitive in our  world  operating on length scales ranging from the atomic scale to cosmological ones. It is at the core of natural life as well as our current technology.  Any, even slight, improvements in transport efficiency can bring profound effects.  In the natural world this can lead to a species dominating another, while in the technological world it can lead to lower energy consumption devices. Many of these improvements in the technological arena have been achieved by better device engineering reducing noise and imperfections. While this seems a perfectly logical approach, recently however it has been found, though nature already knew, that energy transport can be enhanced by adding environmental noise. This counter intuitive behaviour has shed new light on the conventional thinking that noises in transport phenomena should be removed, and it was indeed triggered by energy transport discussions of light harvesting complex of photosynthetic bacteria~\cite{Blankenship}.

The intensive research on four wave mixing in the light harvesting complex, such as the Fenna-Matthews-Olson  (FMO) complex in green sulfur bacteria~\cite{Brixner,Lee,Panitcha}, light harvesting complex in marine algae ~\cite{Collini} or molecular crystals~\cite{Collini10}, has given us clues to understand the mechanisms for efficient quantum energy transport.  The striking long lived coherence observed in these experiments strongly suggests the positive effect of environmental noise.  Extending this concept to the dynamics of exciton motion with delta-correlated stochastic noise (white noise) proposed by  Haken and Strobl~\cite{Haken}, a simple theoretical model using pure dephasing was introduced to assist the excitation transport, referred as the environment-assisted quantum transport (ENAQT)~\cite{Alan} or dephasing-assisted transport~\cite{Plenio}.  
To describe the long-lived coherence observations, the model was further extended to include finite noise correlation times and lengths, including a coloured noise approach using dichotomic stochastic process~\cite{Chen}, a reservoir modeling with an infinite number of quantum harmonic oscillators~\cite{Ishizaki0,Ishizaki1,RCA,MLOR,Plenio15}, and an analysis with finite correlation lengths~\cite{Yu,Fassioli,Cao2009,Bhatta}. 
Positive spatial correlations were shown to reduce the environmental effects, assisting the long-lived coherence. Both effects of temporal and spatial correlation together ~\cite{Wu,Sarovar} were shown to be beneficial, and recently an artificial realization of the ENAQT model has been proposed~\cite{Levi,Bigger}. The long-lived coherence implies the lasting energy exchange between sites, which however may affect the efficiency on the energy transfer rate. If the long-lived coherence can contribute to a higher transport efficiency, these contradicting evidences need to be somehow resolved. Spatial correlation might give us a solution to this~\cite{Cao2009,Nalbach,Tiwari,Fassioli,Bhatta}.  Cao and Silbey~\cite{Cao2009} extended the interaction between the sites to include a phase relation and found the dependency of transfer efficiency on the phase, while the spatial correlation by propagation of environmental phonon modes \cite{Nalbach} and the application of the extended ENAQT model with the spatial correlation to the photosynthetic bacteria~\cite{Fassioli,Bhatta} have been considered.  Anti-correlation in noise has also been taken into account as the effects of anti-correlated (anti-phase) motion of two harmonic oscillators were investigated~\cite{Tiwari}, and the anti-spatial correlations between the bacteriochlorophylls in the FMO complex could both positively and negatively influence the energy transport~\cite{Bhatta}.  
These results imply that the phase relation can affect the quantum energy transport, which leads us to the natural question of what is the best spatio-temporal correlation function.

In this manuscript, we consider a simple multi-site model for the environment-assisted quantum transport using spatio-temporal correlated stochastic noise processes and show how we can engineer the environment to enhance quantum energy transport.   The fundamental difference with the conventional treatment is to exploit the negative parameter regime (anti-correlation) of the spatial correlation as well as including finite temporal correlations with exponential decay.  For the two site model, using the 2nd order of time-convolutionless (TCL) master equation, we analytically find that the time evolution of the population under negative spatial correlation is significantly faster than uncorrelated and positively correlated spatial noise.  Extending the spatial correlations to include anti-correlations provides another control parameter to find efficient energy transfer regimes.  Using two independent measures for transfer efficiency, we find a significant improvement in transported quantity and elapsed time for negative spatial-correlation compared to noise with delta-correlation both in space and time as in the original ENAQT model. Extending to the three site model, we find that the negative spatial correlation between every other site shows the best transfer efficiency, while the negative correlation between the next nearest neighbors is worse. \\

\noindent {\textsf{\bf Results}}\\

\begin{figure}[htb]
\includegraphics[width=9cm]{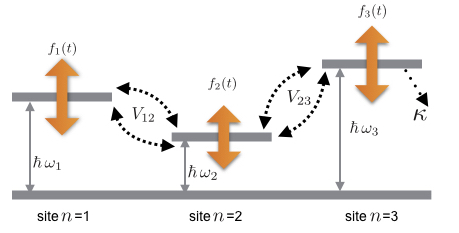}
    \caption{Schematic energy level diagram of our multi-site system illustrates the case of three sites.  Each site $n$ has an associated excited state $|n\rangle$ with Larmor frequency  \(\omega_{n}\), and the excited state at each site fluctuates with the frequency fluctuation \(f_{n}(t)\) around a fixed level.  Each site has a transition frequency \( V_{ij}\) between the adjacent sites \(i=n\) and \(j=n + 1\).  The final site in the chain,  i.e. site n=3 in this illustration, includes a decay channel with rate  \(\kappa\) to trap the system in its ground state. This corresponds to the excitation leaving the chain. }
\label{fig:fig1}
\end{figure}

{\it{Model:}} Let us begin by describing a simple multi-site model for the environment-assisted quantum transport whose energy level diagram we schematically depict in Fig.  (\ref{fig:fig1}).  
For \(N\) sites, the total Hamiltonian can be written as \(\ch(t)=\ch_{0}+  \epsilon \ch_{1}(t)\) with
\begin{eqnarray}
\ch_{0} &=& \hbar \sum_{n=1}^{N}  \omega_{n} |n\rangle \langle n| + \hbar \sum_{n < m} V_{nm} (|m\rangle \langle n|+|n\rangle \langle m|) \label{eqn:1} \\
\ch_{1}(t)&=& \hbar \sum_{n=1}^{N} f_{n}(t) |n\rangle \langle n|, \label{eqn:2} 
\end{eqnarray}
where \(\epsilon\) is the cumulant expansion parameter of the time-convolutionless master equation with \(|n\rangle\) being the \(n\)-th excitation site, \(\omega_{n}\) the \(n\)-th site  Larmor frequency, \(V_{nm}\)  the transition frequency between the \(n\)-th and \(m\)-th site, and \(f_{n}(t)\) the fluctuating frequency on the \(n\)-th site which we consider as a stochastic process. We assume that the average of the frequency fluctuation for each site \(n\) is zero as \(\langle f_{n}(t)\rangle=0\), and the correlation function $\langle f_{n}(0) f_{m}(t) \rangle$ of these fluctuations can for convenience be described by a simple exponential decay as
\begin{eqnarray}
\langle f_{n}(0) f_{m}(t) \rangle  = c_{n,m} \Delta_{n,m}^2 \exp[-|t|/\tau_{c,\{n,m\}}], \label{eqn:11} 
\end{eqnarray}
where to allow for both positive and negative (anti) spatial correlations we introduce the quantity $c_{n,m}$ with \(n,m=\{1,2 \dots N\}\). It is defined over the range by \(-1 \le c_{n,m}  \le 1\) with the extremal values $-1$ ($+1$) corresponding to perfectly anti-correlated (correlated) noise respectively.  \(\Delta_{n,m}\) is the amplitude of the fluctuation, whereas \(\tau_{c,\{n,m\}}\) is the correlation time of the fluctuation.
We also include damping with the rate  \(\kappa\) at the end of the linear chain to  trap the system in its ground state \cite{Cao2009}. This corresponds to the excitation leaving the chain.  Now averaging the density operator over the fluctuation using a time-convolutionless decomposition approach, we obtain the master equation
\begin{eqnarray}
\frac{d}{dt} \rho(t)&=& \langle (-i \el_{0}) \rangle \rho(t)  + \int_{0}^{t} (\langle (-i {\hat \el}_{1}(0)) (-i {\hat \el}_{1} (-\tau)) \rangle - \langle (-i {\hat \el}_{1}(0))\rangle \langle(-i {\hat \el}_{1} (-\tau)) \rangle ) d\tau \rho(t), 
\label{eqn:me}
\end{eqnarray}
where \(\el_{0} X \equiv \frac{1}{\hbar} [\ch_{0}, X]\)  and \(\el_{1}(t) X \equiv \frac{1}{\hbar} [\ch_{1}(t), X]\) for an arbitrary operator \(X\) with \({\hat \el}_{1}(t)=e^{i \el_{0} t} \el_{1}(t) e^{- i \el_{0} t}\).   The master equation given by (\ref{eqn:me}) is inherently non-Markovian and incorporates both spatial and temporal correlations. In the limit that such correlations vanish,  it reduces to a Markovian master equation(see Methods).  Let us now consider a simple example.\\

\begin{figure}[h]
\includegraphics[scale=0.8]{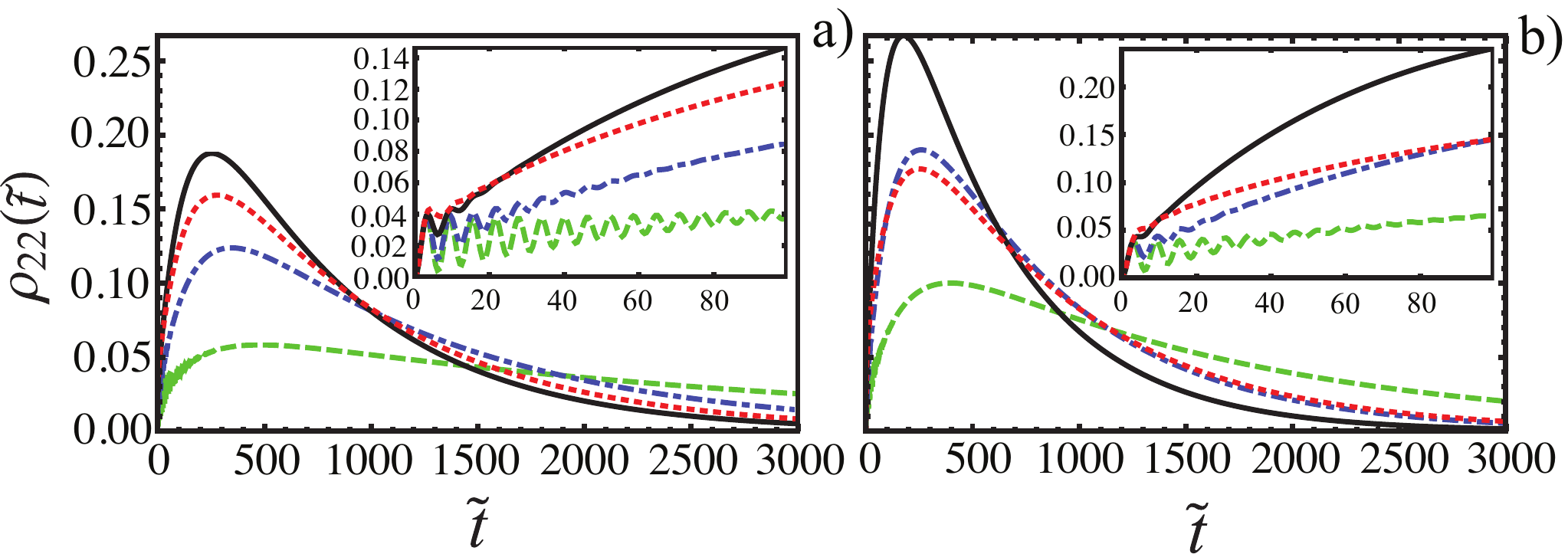}
\caption{Time evolution of \(\rho_{22}({\tilde t}= \Delta \cdot t )\) between energy fluctuation is shown in (a) with spatially-uncorrelated noise, \(c_{n,m}= \delta_{n,m}\) and is shown in (b) with spatially-anti-correlated noise, \(c_{n,m}= -1\), for an initial condition \(\rho_{11}(0)=1\).  Each plots corresponds to a different correlation time,  \(\alpha=\Delta \cdot \tau_{c}\) as \(0.1,0.3,1\) and \(10\): the dashed lines corresponds to \(\alpha=0.1\), dotdashed lines to \(\alpha=0.3\), solid lines to \(\alpha=1\) and dotted lines to \(\alpha=10\).  The system parameters are set as \(\epsilon^2=0.1, \omega_{1}/\Delta=1.5, \omega_{2}/\Delta=0.5,V_{12}/\Delta=0.1 ,\kappa/\Delta (={\tilde \kappa})= 0.005\). The spatially-anti-correlated noise shows the acceleration of energy transport compared with the spatially-uncorrelated case.   }
\label{fig:fig2}
\end{figure}

{\it{ Two-site model}} : As the description of our multi-site model has been completed, we will  now consider the simplest situation,  \(N=2\).  In the two-site model, when initially only the site 1 is fully excited, the time evolution of the probability of finding the second site excited is shown in Fig. (\ref{fig:fig2}). Fig. (\ref{fig:fig2} a) represents the case where the fluctuation has no cross correlation between sites (\(c_{n,m}= \delta_{n,m}\)) with the inset showing the short-time behavior.  The uncorrelated situation (Fig. (\ref{fig:fig2} a)) shows that there is an optimum time to transport the excitation to the second site depending on \(\tau_{c}\), where we set the amplitude and correlation time of fluctuation at each site to be the same as \(\Delta_{n,n}=\Delta\) and \(\tau_{c,\{n,n\}}=\tau_{c}\) for \(n=\{1,2\}\). As the correlation time \(\tau_{c}\) increases, the change of fluctuation carries a longer time memory effect and the optimum time decreases.
Next in Fig. (\ref{fig:fig2} b) we consider the situation where the fluctuation is anti-correlated between sites 1 and 2 by setting \(c_{1,2}=c_{2,1}= -1\).  Assuming that the amplitude and correlation time of the fluctuation of each and between sites are the same,  \(\Delta_{n,m}=\Delta\) and \(\tau_{c,\{n,m\}}=\tau_{c}\)  for \(n,m=\{1,2\}\), we show in Fig. (\ref{fig:fig2} b) that the transport finishes faster as the correlation time becomes longer. Comparing Fig. (\ref{fig:fig2} a) with (\ref{fig:fig2} b), we find that the anti-correlation between the sites makes transport to finish faster than the uncorrelated case. This is not unexpected when one examines the two time noise correlation function given by  \(\phi(t)=\langle(f_{1}(0)-f_{2}(0))(f_{1}(-\tau)-f_{2}(-\tau))\rangle\). For uncorrelated noise this simplifies to  \(\langle f_{1}(0) f_{1}(-\tau)\rangle+\langle f_{2}(0) f_{2}(-\tau))\rangle=2 \langle f_{1}(0) f_{1}(-\tau)\rangle \) if the noise is the same for each site, while for the anti-correlated noise, the quantity becomes, \(\phi(t)=4\langle f_{1}(0) f_{1}(-\tau)\rangle \), to be twice the size of the uncorrelated case.

\begin{figure}[h]
\includegraphics[scale=0.85]{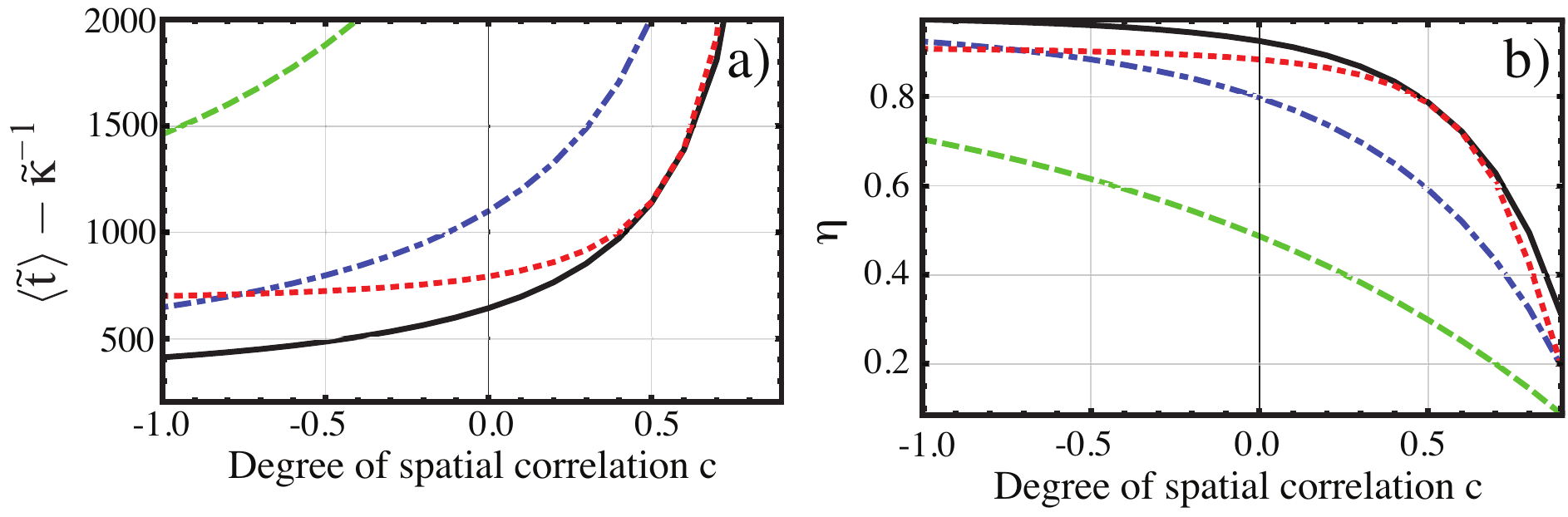}
\caption{Two-site transport properties and their dependence on the degree of spatial correlation. In a), we show the average trapping time \(\langle {\tilde t} \rangle \) minus \({\tilde \kappa}^{-1} \) indicating the average time for the excitation trapped in the system, i.e. in the site 1 for the two-site model.  In b), we show the ratio of transported probability.  In both situations, we find a monotonic dependence on degree of spatial correlation parameter \(c\), which indicates that transport is mostly accelerated for the anti-correlated energy fluctuations.  The dependence of both measures on the correlation time is shown with \(\alpha=\Delta \cdot \tau_{c}\) as \(0.1,0.3,1\) and \(10\). The dashed lines corresponds to \(\alpha=0.1\), dotdashed lines to \(\alpha=0.3\), solid lines to \(\alpha=1\) and dotted lines to \(\alpha=10\).  Focusing on the uncorrelated case, as  \(\alpha\)  increases up to \(1\), the average trapping time (ratio of transported quantity) decreases (increases).  However, for \(\alpha=10\), the former (the latter) increases (decreases), which shows that a suitable correlation time of fluctuation can accelerate the transport.  While the average trapping time (ratio of transported quantity) increases (decreases), for the positive degree of correlation, the acceleration of transport becomes negative. Other parameters are the same as in Fig. (\ref{fig:fig2}).}
\label{fig:fig3}
\end{figure}

To quantify this improvement in more details, let us examine two independent measures: (1) average trapping time and (2) ratio of transported quantity. The average trapping time is defined by \cite{Cao2009}  
\begin{eqnarray}
\langle t \rangle = \sum_{n} \tau_{n} = \sum_{n}\int_{0}^{\infty} dt \rho_{nn}(t). 
\end{eqnarray}
The average trapping time indicates how long the population of the initial excitation remains in the system.  Since the excitation trapping only at the second site with rate \(\kappa\) leads to \(\int_{0}^{\infty} dt\rho_{22}(t)=\kappa^{-1}\), we focus on the average trapping time \(\langle t \rangle \) minus an offset as \(\kappa^{-1} \). This is drawn in Fig. (\ref{fig:fig3} a)  for varying degree of spatial correlation  \(c_{1,2}=c_{2,1}=c\). This indicates that the transport feature depends on the correlation between sites. Especially, as \(c\) approaches to \(-1\), the average trapping time decreases, i.e. the transport finishes faster.  Moreover, the quantum yield becomes larger as c approaches to \(\sim\)1, since the average trapping time is found to be inversely related to the quantum yield \cite{Cao2009} given by \(q \approx {1}/{(k_{d} \langle t \rangle+1)}\) where \(k_{d}\) is the decay rate of recombination (see Methods). 

Next, we introduce a quantity \(\eta\) to define an integrated probability that the excitation has been transported to the second site upto a time \(t_{u}\) as
\begin{eqnarray}
\eta  = \kappa \int_{0}^{t_{u}} \rho_{22}(t') dt'. \label{eqn:13}
\end{eqnarray}
Since the trapping rate \(\kappa\) is given by \(\int_{0}^{\infty} \rho_{22}(t') dt'=\kappa^{-1}\), \(\eta\) indicates the ratio of transported quantity between upto a finite time \(t_{u}\) and completion of the transport.  In Fig.(\ref{fig:fig3} b), we chose \(t_{u}\) to be \(2000 \Delta\) when the transport to the second site is 98\% completed for \(\alpha=1\) and \(c=-1\).  We used it for all other evaluation in the figure. It is straightforward to observe from Figs. (\ref{fig:fig2}) and (\ref{fig:fig3}) the  dependence of the energy transfer on both the degree of spatial correlation \(c\) and  temporal correlation time \(\alpha = \Delta \cdot \tau_c\). In both measure, there is a clear monotonic dependence on the degree of spatial correlation, with the best results occurring as we approach perfect anti-correlation (\(\phi(t)\) is larger in this case).  We also observe a non-monotonic dependence of the populations time evolution and the measures of transfer efficiency for temporal correlations. This indicates that there will be a condition for optimal energy transfer on the correlation time.  For small and decreasing \(\alpha\), the energy transfer takes longer time due to the fact that  fluctuation becomes too fast making it difficult for the energy gaps to be close and the transition probability between sites becomes smaller (this moves us towards the Markovian limit).  For \(\alpha\) larger than the optimal value, the correlation time becomes larger making the transition probability smaller again. Thus for the best performance one should operate near this optimal value.  Now what happens when we have more sites?

\begin{figure}[tb]
\includegraphics[scale=0.8]{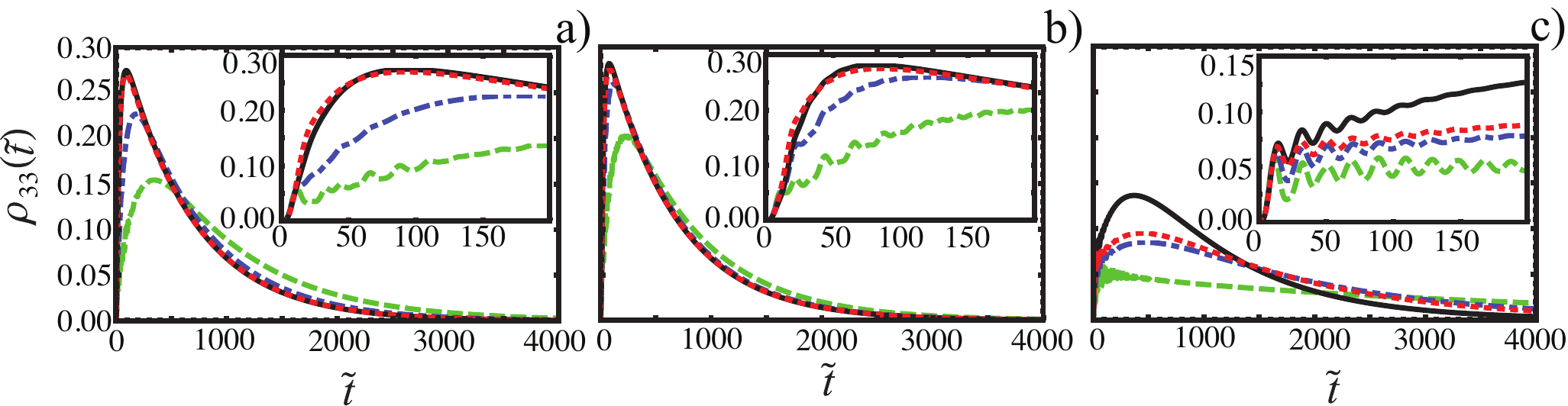}
\caption{Time evolution of \(\rho_{33}({\tilde t}= \Delta \cdot t )\) for the three-site model under different degrees of spatial correlation:  In a), we show the case for energy fluctuations with spatially-uncorrelated noise (\(c_{n,m}= \delta_{n,m}\) for n,m=\{1,2,3\}) while b) shows the anti-correlated between nearest neighbor site case where  \(c_{1,2}= c_{2,1}=c_{2,3}= c_{3,2}=-1\) and positively correlated for site 1 and 3 \(c_{1,3}= c_{3,1}=1\).  c) illustrates the case of anti-correlated between the site 1 and 3, and site 2 and 3, with \(c_{1,3}= c_{3,1}=c_{2,3}= c_{3,2}=-1\) and correlated between the site 1 and 2, with \(c_{1,2}= c_{2,1}=1\).  The dashed lines corresponds to \(\alpha=0.1\), dotdashed lines to \(\alpha=0.3\), solid lines to \(\alpha=1\) and dotted lines to \(\alpha=10\). The inset in each figure shows the short-time behavior.  Comparing the behaviour between the different case, we find that the "anti-ferromagnetic" correlation (case b) accelerates the transport most.  We have set the paramters as \( \omega_{1}/\Delta =1.5 , \omega_{2}/\Delta=1.2, \omega_{3}/\Delta=1.0,V/\Delta=0.1\) and \(\kappa/\Delta (={\tilde \kappa})=0.005\) while other parameters are the same as in Fig. (\ref{fig:fig2}).}
\label{fig:fig4}
\end{figure}

{\it Three-site model.} The two-site model has shown how spatial correlations are potentially an important resource for efficient energy transfer. The natural question which follows to this would be whether this is true for when the system has more than two sites.  We extend the model to three-site linear chain under the nearest neighbor interaction with the interaction strengths given by \(V_{12} = V_{21} = V_{23} = V_{32} = V\) and \(V_{13}= V_{31}=0\).  In Fig. (\ref{fig:fig4}) we illustrate the time evolution of the population of the third (final) site \(\rho_{33}({\tilde t})\) for  spatially-uncorrelated noise and two spatially-anti correlated noise cases. Fig.(\ref{fig:fig4} a) corresponds to the uncorrelated case with \(c_{n,m}= \delta_{n,m}\) for \(n,m=\{1,2,3\}\), while Fig. (\ref{fig:fig4} b) represents the case for the anti-correlated noise between the nearest neighbor sites with \(c_{1,2}= c_{2,1}=c_{2,3}= c_{3,2}=-1\) and correlated between the end sites \(c_{1,3}= c_{3,1}=1\).  Finally, Fig. (\ref{fig:fig4} c) represents the case for the anti-correlated noise between the sites 1 and 3, and the sites 2 and 3, with \(c_{1,3}= c_{3,1}=c_{2,3}= c_{3,2}=-1\) and correlated noise between the sites 1 and 2, \(c_{1,2}= c_{2,1}=1\). In each sub-figure, we find that the dependence of the dynamics on the correlation time for the two-site model remains for the three-site case: transport finishes faster as the correlation time becomes longer particularly in the short correlation time regime, \(\alpha \lesssim 1\). Comparing Fig.(\ref{fig:fig4} a, b, c), we find that the anti-correlated between the nearest neighbor sites typically shows the most efficient transport. 
 
\begin{figure}[h]
\includegraphics[scale=0.4]{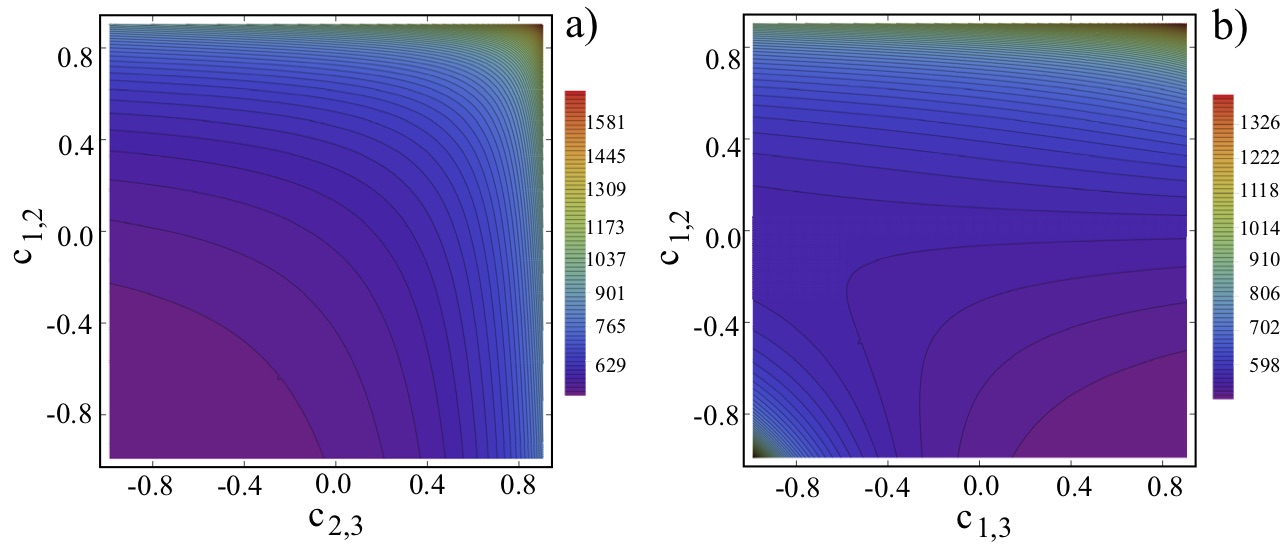}
\caption{Average trapping time \(\langle {\tilde t} \rangle  - {\tilde \kappa}^{-1} \) for three-site transport depending on the degree of spatial correlation for \(\alpha=0.3\).  a) shows  this quantities dependency on\(c_{2,3}(= c_{3,2})\)  and \(c_{1,2}(= c_{2,1})\) while setting \(c_{1,3}= c_{3,1}=c_{2,3} c_{1,2}\). Similarly b) shows  our quantities dependency on \(c_{1,3}(= c_{3,1})\) and \(c_{1,2}(= c_{2,1})\) while setting \(c_{2,3}= c_{3,2}=c_{1,3} c_{1,2}\). Other parameters are the same as in Fig. (\ref{fig:fig4}). The shorter values of average trapping time occurs at the left hand bottom side corner in a) and right hand bottom side corner in b). This means the ``anti-ferromagnetic" correlation case is the best choice for acceleration of transport in this case. }
\label{fig:fig5}
\end{figure}

To explore these features systematically, we evaluate the average trapping time \(\langle t \rangle \). In Fig.(\ref{fig:fig5}), we show the contour plot of the dependence of \(\langle t \rangle -\kappa^{-1}\) on the degree of correlation \(c_{n,m}\). 
Comparing Figs.(\ref{fig:fig5} a,b), we find that ``anti-ferromagnetic" correlation where the fluctuation is anti-correlated between the adjoint site shows the most efficient transport.  The origin of the efficient transport lies at the higher transition probability between the adjoint sites with anti-correlation, which in the three site model corresponds to the``anti-ferromagnetic" correlation.  There the average trapping time \(\langle t \rangle  -\kappa^{-1} \) for the uncorrelated case is $\sim12$ \% longer than the ``anti-ferromagnetic" case. This clearly shows the acceleration in the energy transport occurs with ``anti-ferromagnetic" correlations, while the difference is smaller as \(\alpha\) increases (5 \% for \(\alpha=0.5\)).  The question follows is how robust this efficiency in energy transport.
 
 The most important feature is the stability against inhomogeneity in coupling between adjoint sites.  To see this, we deviate \(V_{23}/\Delta\) from \(0.1\) to be \(0.15\) and \(0.3\), keeping  \(V_{12}/\Delta =0.1\) in the three site model.  The average trapping time decreases for any spatial correlation as increasing \(V_{23}/\Delta\), due to the stronger coupling strength accelerating the energy transfer. 
The difference of the average trapping time \(\langle t \rangle \) between uncorrelated and ``anti-ferromagnetic" case decreases from  $\sim12$ \%  to  $\sim8.5$ \% for \(V_{23}/\Delta=0.15\) and $\sim4$ \% for \(V_{23}/\Delta=0.3\).   This indicates a significant robustness in the acceleration mechanism, yielding the robustness around $\sim10$ \% with the inhomogeneous coupling upto 150\%. \\
\\

\noindent {\textsf{\bf Discussion and Conclusion}}\\
Energy transport in molecular complexes / light harvesting complexes in photosynthetic bacteria has raised many fundamental questions on how our nature operates and  our understanding of noise effects in such processes has been challenged. The recent research indicates that noise can enhance transport rates through temporal  correlations \cite{Ishizaki0,Ishizaki1,RCA,MLOR,Plenio15} or spatial correlations \cite{Fassioli}. In this work we have considered the transport of excitation for a multi-site linear chain model whose energy levels are affected by spatio-temporal correlated stochastic noise processes \cite{Yu,Cao2009}. We find that the energy transport can be accelerated by extending the spatial correlation into the negative region (anti-correlation).  In the two-site model, our numerical analysis showed the significant acceleration in the energy transfer with the negative spatial correlation.  By extending the model to three sites, we numerically demonstrated that the optimal efficiencies can be obtained with the ``anti-ferromagnetic" correlation.  The difference in transport with anti-correlated and uncorrelated noises in the four site model becomes even larger.
For the temporal correlation dependencies, we explored the energy transfer dependence by changing the parameter  \(\alpha= \Delta \cdot \tau_{c}\).  A non-monotonic dependence on both the population time evolution and the transfer efficiency measures were observed, and hence the correlation time needs to be chosen to achieve the optimum energy transport. These results show new possibilities to understand efficient energy transport in nature and engineer it to our technologies.\\

\vskip 1truecm
\noindent {\textsf{\bf Methods}}\\

{\it An open systems approach:} As shown in Fig (\ref{fig:fig1}) our multi-site model contains  fluctuating excited  state energy levels as well as an energy trap on the last site (energy decay). In principle this means an open systems approach must be used - especially as our energy level fluctuations are stochastic in nature. It would thus be natural to write down a master equation in Lindblad form  \cite{GKS,Lindblad} (which can easily handle the last site energy decay), which would force us to use a white noise model where the fluctuations are not correlated in time. However in this case we want to examine temporal and spatial correlation effects. This means the typical master equation is not appropiate, however a master equation can be derived using time-convolutionless decomposition techniques~\cite{Kubo,Hanggi,HSS,KTH,STH,CS,FA,US,Breuer}.\\
 
 {\it Derivation of the time convolution type of master equation:}\\
The master equation given by Eq.(4) is obtained by extracting the time evolution of the excitations in each site from the one of the total system which is written by the Liouville-von Neuman equation as
\begin{eqnarray}
\frac{d}{dt} W(t)=-i \el(t) W(t), \label{eqn:M1}
\end{eqnarray}
where \(W(t)\) is the density operator for the total system and  \(\el(t)\) is the Liouville operator defined as \(\el(t) X =\frac{1}{\hbar} [\ch(t), X] \) for an arbitrary operator \(X\).
In such a case, the extraction averages \(W(t)\) over the stochastic process of the fluctuation. Our purpose is to obtain the time evolution of the reduced density operator \(\langle W(t)\rangle (\equiv \rho(t)) \) where we denote the average operation as \( \langle \cdots \rangle\).  For this purpose, it is convenient to use the projection operator method \cite{Kubo,Hanggi,HSS,KTH,STH,CS,FA,US,Breuer}.  Introducing a projection operator, \(\cp\), which is an idempotent operator with a property as \(\cp^2=\cp\), we describe the reduced density operator as  \(\cp W(t) \equiv \langle W(t) \rangle (=\rho(t))\).   To obtain the time evolution of \(\cp W(t)\), we use the formal solution of Eq.~(\ref{eqn:M1}) as \(W(t)=U_{+}(t,t_{0}) W(t_{0})\) with \(U_{+}(t,t_{0})=T_{+} \exp[-i \int_{t_{0}}^{t} \el(\tau) d\tau]\) where \(T_{+}\) is an increasing time ordering operator from the right to the left.  The procedure to obtain the master equation, Eq.(4), is roughly divided into the following six steps:
\begin{enumerate}
\item First we define the relevant \(\cp\) part and the irrelevant \(\cq (\equiv 1-\cp)\) part of the time evolution operator \({\hat U}_{+}(t,t_{0})\) as 
\begin{equation}
x(t) \equiv  \cp {\hat U}_{+}(t,t_{0}), \;\;\;  y(t)\equiv \cq {\hat U}_{+}(t,t_{0}), \label{eqn:M2}
\end{equation} 
where we use interaction picture with the definition as \({\hat U}_{+}(t,t_{0})= e^{-i \el_{0}(t-t_{0})} U_{+}(t,t_{0})=T_{+} \exp[-\int_{t_{0}}^{t} dt' i {\hat \el}_{1}(t') dt'] \) and \({\hat \el}_{1}(t)= e^{i \el_{0}(t-t_{0})}\el_{1}(t)e^{-i \el_{0}(t-t_{0})}\).

\item Then we derive simultaneous differential equations for  \(x(t)\) and \(y(t)\) as
\begin{eqnarray}
\frac{d}{dt} x(t)&=&\cp (- i {\hat \el}_{1}(t)) x(t) + \cp (- i {\hat \el}_{1}(t)) y(t) \; , \label{eqn:M3}\\
\frac{d}{dt} y(t)&=&\cq (- i {\hat \el}_{1}(t)) x(t) + \cq (- i {\hat \el}_{1}(t)) y(t) \; .\label{eqn:M4}
\end{eqnarray}

\item Next the formal solution of the irrelevant \(\cq\) part can be written as,
\begin{eqnarray}
y(t)&=&\int_{t_{0}}^{t} {\hat \cv}_{+}(t,\tau) \cq (-i {\hat \el}_{1}(\tau)) x(\tau) d\tau + {\hat \cv}_{+}(t,t_{0}) \cq.  \label{eqn:M5}
\end{eqnarray}
with \({\hat \cv}_{+}(t,\tau)=T_{+} \exp[- i \int_{\tau}^{t} \cq {\hat \el}_{1} (\tau') d\tau']\).

\item We then rewrite \(x(\tau)\) in Eq.~(\ref{eqn:M5}) with \(x(t)\) and \(y(t)\)  using the relation 
\begin{equation}
x(\tau)=\cp {\hat U}_{+}(\tau,t_{0}) =\cp {\hat U}_{-}(t,\tau) (x(t)+y(t)), \label{eqn:M6}
\end{equation}
where \({\hat U}_{-}(t,t_{0})=T_{-} \exp[i \int_{t_{0}}^{t} {\hat \el}_{1}(\tau') d\tau']\) with \(T_{-}\) an increasing time ordering operator from the left to the right. The formal solution of \(y(t)\) has the form
\begin{equation}
y(t)= \theta(t)^{-1}( (1-\theta(t)) x(t)+{\hat \cv}_{+}(t,t_{0}) \cq), \label{eqn:M7}
\end{equation}
where we define
\begin{equation}
\theta(t)=1- \int_{t_{0}}^{t} {\hat \cv}_{+}(t,\tau) \cq (-i {\hat \el}_{1}(\tau)) \cp {\hat U}_{-}(t,\tau) d\tau  \equiv 1-\sigma(t). \label{eqn:M8}
\end{equation}

\item Next we substitute the formal solution of \(y(t)\) into Eq.~(\ref{eqn:M3}) and multiply \(W(t_{0})\) from the right on the both hand sides of Eq.~(\ref{eqn:M3}),
\begin{equation}
\frac{d}{dt} {\hat \rho}(t)= \cp (-i {\hat \el}_{1}(t)) {\hat \rho}(t)+ \Xi(t,t_{0}) {\hat \rho}(t) + I (t,t_{0}) W(t_{0}), \label{eqn:M9}
\end{equation}
with \({\hat \rho}(t)=e^{-i \el_{0}(t-t_{0})}  \rho(t)\), \(\Xi(t,t_{0})=\cp (-i {\hat \el}_{1}(t)) (1-\theta(t)^{-1}) \) and \(I(t,t_{0})=\cp (-i {\hat \el}_{1}(t)) \theta(t)^{-1} {\hat \cv}(t,t_{0}) \cq\).

\item Eq.~(\ref{eqn:M9}) is then expanded with using the relation as \(\theta(t)^{-1}=\sum_{n=0}^{\infty} \sigma(t)^{n}\). This gives
\begin{equation}
\frac{d}{dt} {\hat \rho}(t)= \cp (-i {\hat \el}_{1}(t)) {\hat \rho}(t)+ \cp (-i {\hat \el}_{1}(t))\sum_{n=1}^{\infty} \sigma(t)^{n} {\hat \rho}(t) + I (t,t_{0}) W(t_{0}), \label{eqn:M10}
\end{equation}
\item Expansion of Eq.~(\ref{eqn:M10}) up to the second order of cumulant for \(\el_{1}(t)\) with using an assumption as \(\langle f_{m}(t)\rangle=0\) and \(\cq W(t_{0})=0\) and transformation into the original picture from the interaction picture gives Eq.(4).  
\end{enumerate}
Now let us consider the specific example, two site model.\\

{\it TCL type master equation for the 2-site model:}\\
The concrete form of the master equation for the two-site model is written as
\begin{eqnarray}
\frac{d}{dt} \rho_{11}(t) &=& - iV_{12} (- \rho_{12}(t)+\rho_{21}(t)), \nonumber \\ 
\frac{d}{dt} \rho_{12}(t) &=& - i \{V_{12} (- \rho_{11}(t)+\rho_{22}(t))+(\omega_{1} -\omega_{2}) \rho_{12}(t))\}- \epsilon^2  \{ F_{1}(\omega_{1}, \omega_{2},V_{12},t) (\rho_{11}(t)-\rho_{22}(t))+F_{2}(\omega_{1}, \omega_{2},V_{12},t) \rho_{12}(t) \}, \nonumber \\ 
\frac{d}{dt} \rho_{21}(t) &=& - i \{V_{12} (\rho_{11}(t)-\rho_{22}(t))-(\omega_{1} -\omega_{2}) \rho_{21}(t))\}-\epsilon^2  \{ F_{1}^{*}(\omega_{1}, \omega_{2},V_{12},t) (\rho_{11}(t)-\rho_{22}(t))+F_{2}(\omega_{1}, \omega_{2},V_{12},t) \rho_{21}(t) \} , \nonumber \\  
\frac{d}{dt} \rho_{22}(t) &=& - iV_{12} ( \rho_{12}(t)-\rho_{21}(t))-\kappa \rho_{22}(t) , \nonumber 
\end{eqnarray}
where \(\rho_{nm}(t)\) is the \((n,m)\) element of the reduced density operator \(\rho(t)\), \(\kappa\) is the trap frequency at the 2nd site and we \(F_{n}(\omega_{1}, \omega_{2},V_{12},t)\) for \(n=1,2\) are defined as
\begin{eqnarray}
F_{1}(\omega_{1}, \omega_{2},V_{12}, t )&=& - \int_{0}^{t} d{\tau} \Bigl( \frac{V_{12}(\omega_{1}-\omega_{2})}{\mu^2}(1-\cos{(\mu \tau)})-i \frac{V_{12}}{\mu} \sin{(\mu \tau})\Bigr) \phi(\tau), \label{eqn:7} \\
F_{2}(\omega_{1}, \omega_{2},V_{12}, t )&=& \int_{0}^{t} d{\tau}  \Bigl((\frac{(\omega_{1}-\omega_{2})}{\mu})^2+(1-(\frac{(\omega_{1}-\omega_{2})}{\mu})^2)\cos{(\mu \tau)}\Bigr) \phi(\tau), \label{eqn:8} 
\end{eqnarray}
with
\begin{eqnarray}
\mu&=&\sqrt{(\omega_{1}-\omega_{2})^2+ 4V_{12}^2}, \label{eqn:9} \\
\phi(t)&=&\langle (f_{1}(0) - f_{2}(0))(f_{1}(-t)-f_{2}(-t))\rangle.  \label{eqn:10} 
\end{eqnarray}
In Eq.~(\ref{eqn:10}), \(\langle f_{n}(0) f_{m}(-\tau) \rangle \) with \(n,m=\{1,2\}\) is the correlation function of the fluctuation of the frequency.
The obtained master equation shows that the transport is controlled by adjusting correlation function of the fluctuation on each site as well as the one between the fluctuation on the different cite such as \(\langle f_{1}(0)f_{2}(-\tau)\rangle\).  Assuming that the amplitude and correlation time of the fluctuation of each and between site are the same,  \(\Delta_{n,m}=\Delta\) and \(\tau_{c,\{n,m\}}=\tau_{c}\)  for \(n,m=\{1,2\}\), \(\phi(t)\) in Eq.~(\ref{eqn:10}) is given by \(\phi(t)= 4 \langle f_{1}(0) f_{1}(-t) \rangle\) for anti-correlated case (\(c=-1\)) and \(\phi(t)= 2 \langle f_{1}(0) f_{1}(-t) \rangle\) for  the uncorrelated case (\(c=0\)). \\ \\
The dynamics described by time convolutionless(TCL) type of master equation is compared to the one by hierarchical equations of motion (HEOM) for the spin-boson system in \cite{FLG} where they find that the second order and fourth order of TCL equation and HEOM shows almost the same dynamics for weak coupling case.\\

{\it Relation with the traditional Lindblad master equation:}\\
The TCL master equation given by Eq.(4) in a specific limit reduces to the typical Lindblad master equation. In such a case we set the upper integral limit $t\rightarrow \infty$ giving 
\begin{eqnarray}
\frac{d}{dt} \rho(t)&=& \langle (-i \el_{0}) \rangle \rho(t)  + \int_{0}^{\infty} (\langle (-i {\hat \el}_{1}(0)) (-i {\hat \el}_{1} (-\tau)) \rangle - \langle (-i {\hat \el}_{1}(0))\rangle \langle(-i {\hat \el}_{1} (-\tau)) \rangle ) d\tau  \rho(t), 
\end{eqnarray}
where all of the coefficients are time-independent. This is the Born-Markov approximation \cite{Breuer,HSS,KTH}. 
For example, the time-dependent coefficients of the two site model ignoring spatial correlation (\(c=0\)) can be approximated as
\begin{eqnarray}
F_{1}(\omega_{1}, \omega_{2},V_{12}, \infty )&=& 
- 2 \Delta^2 \Bigl( \frac{V_{12}(\omega_{1}-\omega_{2}) \tau_{c}^3}{1+\mu^2 \tau_{c}^2}-i \frac{V_{12}\tau_{c}^2}{\mu^2 \tau_{c}^2+1}\Bigr)  \nonumber \\
F_{2}(\omega_{1}, \omega_{2},V_{12}, \infty)&=& 
 2 \Delta^2 \Bigl(\frac{\tau_{c}}{1+\mu^2 \tau_{c}^2}+(\frac{(\omega_{1}-\omega_{2}) }{\mu})^2\frac{\mu^2 \tau_{c}^3}{1+\mu^2 \tau_{c}^2}\Bigr)  ,  \nonumber \\
\end{eqnarray}
Taking the limit of the correlation time \(\tau_{c}\) to approach zero with maintaining \(\Delta ^2 \tau_{c}\) finite \cite{KTH}, we obtain
\begin{eqnarray}
\lim_{\tau_{c} \to 0} F_{1}(\omega_{1}, \omega_{2},V_{12}, \infty ) =0, \;\;\;\;
\lim_{\tau_{c} \to 0} F_{2}(\omega_{1}, \omega_{2},V_{12}, \infty ) =2 \frac{\Delta^2 \tau_{c}}{1+\mu^2 \tau_{c}^2}, 
\end{eqnarray}
Defining \(\lim_{\tau_{c} \to 0} F_{2}(\omega_{1}, \omega_{2},\Gamma_{12}, \infty ) \equiv \gamma \), our master equation given by Eq.(4) reduces to the Lindblad form \cite{GKS,Lindblad} :
\begin{eqnarray}
\frac{d}{dt} \rho(t)=-\frac{i}{\hbar}[\ch_{0},\rho(t)] + \frac{ \gamma}{2}  \sum_{m} [ 2 A_{m} \rho(t) A_{m}^{\dagger} - A_{m}^{\dagger} A_{m} \rho(t)- \rho(t) A_{m}^{\dagger} A_{m}   ], \label{eqn:24} 
\end{eqnarray}
where \(A_{m}=| m \rangle \langle m |\).  We find that Eq.~(\ref{eqn:24}) is the same as the master equation obtained in \cite{Alan} for the Haken-Strobl model. Moreover, when we substitute the stationary value of \(\rho_{12}(t)\) into the differential equation, we obtain the same differential equation as in~\cite{Cao2009}.  Besides the above, let us note that, by using the time convolutionless type of master equation, we need only to take long-time limit as the approximation procedure, which is much simpler than using the time-nonlocal (time-convolution) type of master equation \cite{HSS}.

{\it Numerical methods:} \\
The effect of the fluctuation on the time evolution of the density operator is described with the time dependent coefficient of the second term in the right hand side of the time-convolutionless type of master equation given by Eq. (4).    To obtain the time evolution of the density operator, we numerically solved the master equation by evaluating the time dependent coefficient and iterating the equation step by step with the coefficient.   In the evaluation, we scaled the time variable and parameters with the strength of fluctuation \(\Delta\).  \\

 {\it Quantum yield:}\\
The quantum yield  is defined as
\begin{eqnarray}
q=\frac{\sum_{n} k_{t,n} \tau_{n}}{\sum_{n} k_{t,n} \tau_{n}+\sum_{n} k_{d,n} \tau_{n}},  \label{eqn:25} 
\end{eqnarray}
where  \(k_{d,n}\) is the the decay rate at the \(n\)-th site by recombination, \(k_{t,n}\) is the one by trap and \(\tau_{n}=\int_{0}^{\infty} \rho_{nn}(t) dt\) in [17]. \(q\) indicates the ratio of trapped quantity to the total loss by recombination and trap.
Cao and Silbey showed that the decay rate by recombination is necessary to be much smaller than the trapping rate, \(k_{d,n} \ll k_{t,n}\) to obtain a high quantum yield[17].  In such situation,  the dependence of \(\tau_{n}\) on \(k_{d,n}\) can be neglected to give \(\sum_{n} k_{t,n} \tau_{n} (k_{d,n}=0) \approx 1\). Thus
\begin{eqnarray}
q &\approx& \frac{1}{1+\sum_{n} k_{d,n} \tau_{n}(k_{d,n}=0)} = \frac{1}{1+k_{d} \langle t \rangle}, \label{eqn:26} 
\end{eqnarray}
where the last form is obtained by setting the values of \(k_{d,n}\) to be the same as \(k_{d}\) for all of the state \(n\) and with \(\langle t \rangle=\sum_{n} \tau_{n}\), which is called as the average trapping time. \\

{\it Data availability:}\\
The data that support the findings of this study are available from the corresponding author upon reasonable request. \\ 

\noindent {\textsf{\bf Aknowledgements}}\\
We would like to thank Ryuta Tezuka, Ryuji Nakamura, Shun Imazawa, and Dr. Kazunari Hashimoto for valuable discussions. This work was supported in part from the MEXT KAKENHI Grant-in-Aid for Scientific Research on Innovative Areas Science of hybrid quantum systems Grant No.15H05870, JSPS KAKENHI Grant Numbers 15K05200, and the researcher exchange promotion program of the Research Organization of Information and Systems. This project was also made possible through the support of a grant from the John Templeton Foundation. The opinions expressed in this publication are those of the authors and do not necessarily reflect the views of the John Templeton Foundation (JTF \#60478). 
\\

\noindent{\textsf{ \bf Competing Interests}}\\ 
The authors declare no conflict of interest.\\

\noindent{\textsf{ \bf Author Contributions}}\\ 
All authors researched, collated, and wrote this paper.\\
\\

\end{document}